# Research on the construction method of vehicle driving cycle based on Mean Shift clustering

Yongjiang He

**ABSTRACT:** In this study, a novel method for the construction of a driving cycle based on Mean Shift clustering is proposed to solve the problems existing in the traditional micro-trips method. Firstly, 1701 kinematic segments are obtained by processing and dividing the driving data in real road conditions. Secondly, 12 kinematic parameters are calculated for each segment, and the dimensionality of parameters is reduced through principal component analysis (PCA). Three principal components are chosen to classify all cycles into three types by the Mean Shift algorithm. Finally, according to the principle of minimum deviation, representative micro-trips are selected from each type of cycle to complete the construction of the final driving cycle. Further, the construction method in this paper is compared with the micro-trips construction method by the K-Means clustering. The results show that the construction method by Mean Shift clustering can more effectively reflect the real driving data. This study realizes the innovation in the construction method of micro-trips and provides a preliminary theoretical basis for the formulation of automobile working condition standards, energy management of new-energy vehicles, and optimal control of vehicle dynamics in driverless vehicles.

## 1. Introduction

The study of the vehicle driving cycle has always been basic research in the field of automobiles. Through the multivariate statistical analysis of vehicle driving data in the road, we can obtain the time and speed curve that can reflect the typical driving condition of vehicles. The fuel economy analysis and pollution emission of traditional vehicles [1, 2] the control strategy of hybrid electric vehicles [3], and the energy management of electric vehicles [4] are all related to the construction of vehicles driving cycles. In the future, driverless cars will become the development direction of the automobile industry, and their energy management and control strategies also depend on the study of driving cycles. Therefore, a reliable and effective construction method of driving cycles is particularly necessary.

At present, the research on the driving cycle construction method is mainly divided into two construction methods which are based on real driving data [5] and software simulation [6]. Among them, driving cycle construction based on real driving data is the most widely used, especially the micro-trips method [7, 8] and Markov chain method[9, 10]. In this paper, we mainly focus on the micro-trips method, including the division of kinematic segments, cycle classification; and driving cycle synthesis. Where in the kinematic segment division process, a continuous trip will be divided into several cycles according to the definition of kinematic, and cycle classification is to classify each segment according to its kinematic characteristics and then analyze the characteristics of each type of cycle. Driving cycle synthesis is to select representative cycles from each category and combine them into the final conditions. Therefore, in the driving cycle construction of the micro-trips method the classification of segments always has the most significance and has most study.

The classification of kinematic segments based on its kinematic parameters, which usually involve more than a dozen indicators, and it would cause a high-dimensional space. If all parameters are taken as variables to construct driving cycles that bound to complicate the study and even mistakes, so it is necessary to reduce the dimension of parameters [11]. Principal component analysis [12], as a statistical method, is often used in parameter dimensionality reduction and has been widely used in the construction of micro-trips method [13-15]. Besides, Qin Shi et al. [16] also proved the effectiveness of this method in the construction of the driving cycle.

In the micro-trips method, segment classification is realized mainly through the clustering algorithm, among which K-Means clustering is the algorithm commonly used in the current research. Fotouhi et al. [17] divided the kinematic segments of driving data in Tehran and realized the construction of the driving cycle in the city based on the K-Means clustering algorithm. Daniel Förster [18] applied the K-Means algorithm to classify kinematic segments' eigenvalues and then selected each group of eigenvalues by genetic mixed-integer optimization algorithm to complete the construction of a representative driving cycle. Zheng Chen et al. [19], based on the classification of segments by the K-Means clustering method, then selected the characteristic parameters of each group cycle through the convolutional neural network to realize the construction of the driving cycle. On the basis of K-Means clustering, Xuan Zhao et al. [20] added the support vector machine to improve the classifying effect of kinematic segments and proved that the driving cycle established by this method could better reflect the driving condition of the real world.

It can be seen from the above analysis that K-Means, as a common clustering algorithm, is widely used in running fragment classification, and has been improved by using it with other methods. However, the algorithm itself has certain defects. First of all, the K-Means algorithm needs to define the number of clustering in advance, which makes the clustering result with certain subjective factors. Emre Celebi [21] pointed out that it is sensitive to the initial value (clustering centers) when using the K-Means in discrete data. If the initial value is selected at the edge or isolated point, it may lead to confusion of clustering results. Also, Nayini et al. [22] funded that the K-Means clustering algorithm is based on the minimum square error so that to lead to the local optimum. Besides, in the data space, points in the low-density region are often considered as noise [23], and another disadvantage of K-Means is that it cannot identify noise points but divides all points into a predefined category, which increases the uncertainty of clustering results.

In contrast, the Mean Shift clustering method can well avoid the above problems. Hyrien Ollivier et al. [24] pointed out that Mean Shift is a clustering algorithm that simulates the movement of data points, and this it does not assume the number of clusters but relies on the density distribution characteristics of points to realize clustering analysis. More importantly, Shiu SY et al. [25] proved that when there are noise points in the data, this algorithm can isolate noise data points and conduct clustering at the same time. Wang Xiaogang et al. [26] studied the theoretical convergence of the Mean Shift clustering method with large data sets and proved that the convergence of the algorithm depends on its ability to transform the data points into multivariate normal distribution data. Ryoya and Toshiyuki[27] further proved the convergence of the Mean Shift algorithm in the case of using kernel functions.

Compared with the K-Means, the Mean Shift algorithm has obvious advantages in both the clustering principle and noise processing. However, at this stage, in the study of driving cycle construction with micro-trips method has failed to apply the Mean Shift clustering algorithm. Therefore, in this paper, we propose a vehicle driving condition construction scheme based on the Mean Shift clustering algorithm, which can solve the problem of initial value sensitivity and noise under the condition of discrete data to make the driving cycle more suitable for the driving characteristics in the real road.

Nowadays, new energy technologies are driving a novel round of revolution in the automobile industry. Energy management and vehicle dynamics control have gradually become one of the hot topics in unmanned driving research, among which the formulation of operating condition standard plays a key role. Therefore, against the problem existing in driving cycle construction methods, this paper proposes a more reliable method, which can improve the accuracy and reliability of driving cycle construction and provide a research basis for energy management and power control of new-energy vehicles and driverless vehicles.

## 2. Construction scheme design and symbol definition

The construction scheme and implementation of the driving cycle in this paper are shown in **Fig. 1** and the research process is based on the micro-trips method. First, we acquire the vehicle running data in the real road and handling abnormal data. Second, divide a continuous trip into several cycles according to the definition of micro-trips and calculate the characteristic parameters of each segment. Then the dimensionality of feature parameters is reduced by the PCA, and the cycles are classified by Mean Shift with the selected principal components. Finally, the segments used to construct the final driving cycle are constructed by selecting the minimum characteristic deviation of each type, and the construction effect is further analyzed. **Table** 1 shows the symbol definitions and units involved in the description of kinematic characteristics in this paper.

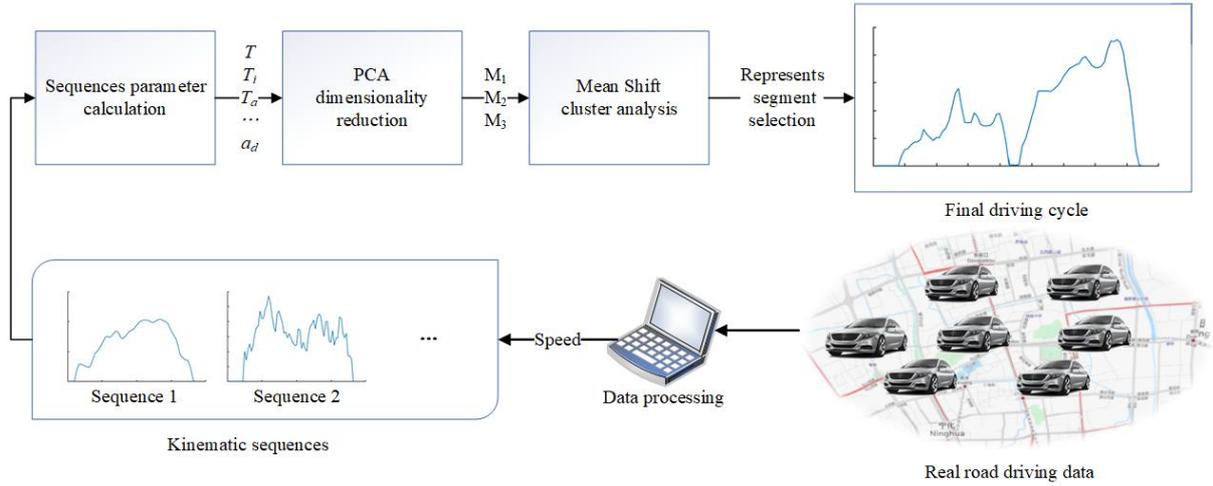

**Fig. 1.** Design of the vehicle driving cycle construction in this paper

**Table 1**

Symbol definitions and the description of kinematics features in this paper

| Number | Symbol | Description | Unit |
| --- | --- | --- | --- |
| 1 | $T$ | Driving time | s |
| 2 | $T_i$ | Idle time | s |
| 3 | $T_a$ | Acceleration time | s |
| 4 | $T_d$ | Decelerate time | s |
| 5 | $T_e$ | Uniform time | s |
| 6 | $S$ | Driving distance | km |
| 7 | $V_{max}$ | Maximum speed | km/h |
| 8 | $V_m$ | Average speed | km/h |
| 9 | $V_{mr}$ | Average travel speed | km/h |
| 10 | $V_s$ | Speed standard deviation | km/h |
| 11 | $a_a$ | Average acceleration | m/s$^2$ |
| 12 | $a_d$ | Average decelerate | m/s$^2$ |

## 3. Data processing

*3.1. Data sources*

The experimental data are the driving data of light cars in Fuzhou, China, includes the driving characteristics of urban roads, rural roads, and expressways of various grades. The data collection time interval is 1 second, including the speed, latitude and longitude, instantaneous fuel consumption, engine speed, and other data. Data collecting over three time periods, respectively are 2017.11.1-

2017.11.7, a total of 145,826 records; 2017.12.1-2017.12.6, a total of 185,726 records; 2017.12.18-2017.12.24 a total of 164,951 records. The data collecting periods all include working days and weekends and any time slots of a day but mainly concentrate from 8:00 to 21:00. To make our research more generally, holidays are not included in the data acquisition period.

*3.2. Data processing*

*3.2.1. GPS abnormal data processing*

**Fig. 2** shows the GPS data of vehicles running from 017.12.1-2017.12.6. The abnormal data of GPS are mainly shown as data beyond the research range and. Among them, the longitude and latitude of the data beyond the research range are all 0°, which can directly eliminate. Also, during the data collecting process, data will get lost due to the shielding of buildings or equipment broken, resulting in data interrupt, as shown in the blue circle in **Fig. 2**. The discontinuous data caused by data interrupt will be removed.

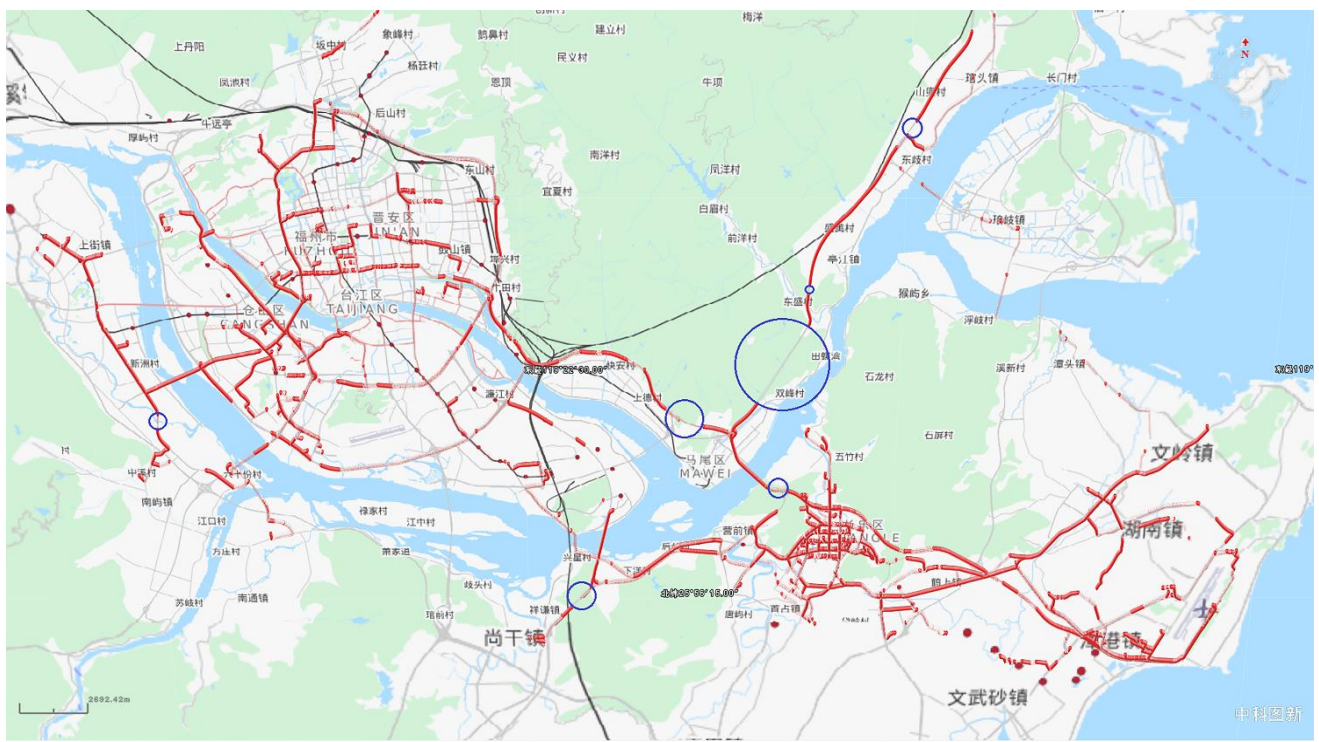

**Fig. 2.** GPS interrupt data in the research area

*3.2.2. Acceleration and deceleration outlier data processing*

The test vehicles are light cars. According to the general performance of a light car, the acceleration is generally less than 4.5m/s², and the deceleration more than -7.5m/s²[28]. For these abnormal data, we adopt the method of averaging the speed before and after. If the acceleration of two adjacent records is higher than 4.5m/s² or the deceleration less than -7.5 m/s², the correction method is as:

$$v_t = (v_{t+1} + v_{t-1})/2 \tag{1}$$

where $v_t$ is the velocity to be corrected, $v_{t-1}$ is the velocity at the previous point, and $v_{t+1}$ is the velocity at the next point.

**Fig. 3** is the correction result of 164 acceleration and deceleration abnormal data from 2017.12.1-2017.12.6, in which the blue points represent the abnormal acceleration before correction, and the red points represent the modified data. The modified acceleration conforms to the general performance of the experimental vehicle.

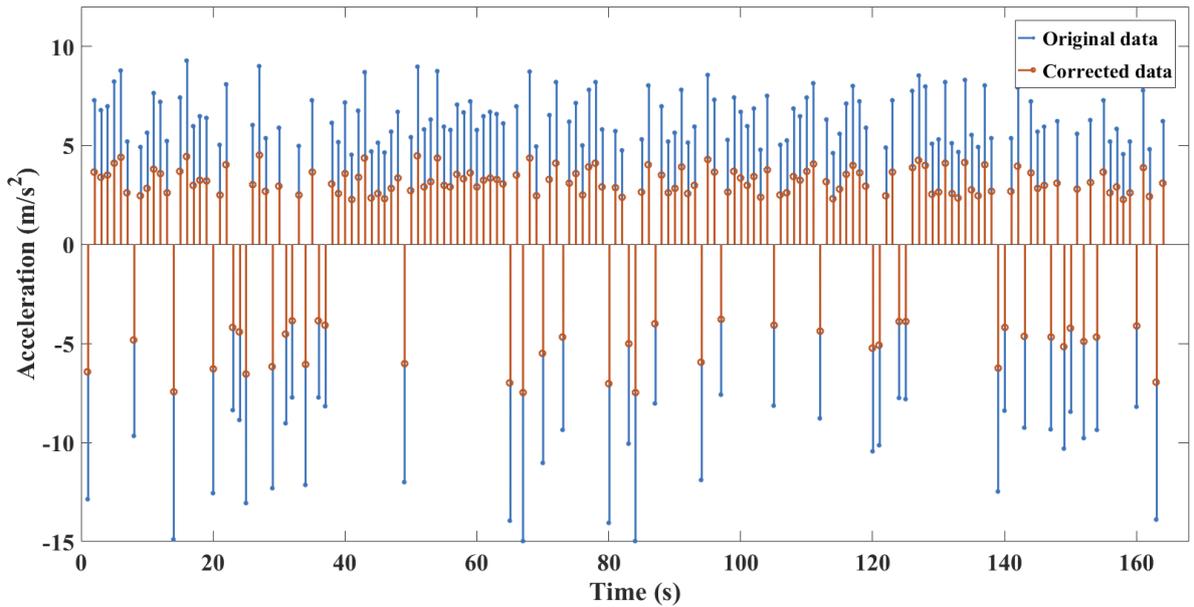

**Fig. 3.** Correction of abnormal data in acceleration and deceleration

*3.2.3. Travel time anomaly data processing*

The abnormal travel time includes a short travel time, unusual idle time, and long parking time [29]. Among them, a short travel time refers to the travel time less than 10 seconds, and the travel speed is less than 10 km/h. In the driving cycle, it is shown as the isolated point or burr data of the curve. As shown in **Fig. 4**, the burr point in the curve can be set to zero. Abnormal idle time refers to the idle time over 180 seconds, and longtime parking refers to the data with a speed of zero exceeding 300 seconds. When dealing with the exception idle time, we pick the last 180 seconds as real idle time. The construction of the vehicle driving cycle should not include the data of parking mode, so the long parking time data should be deleted.

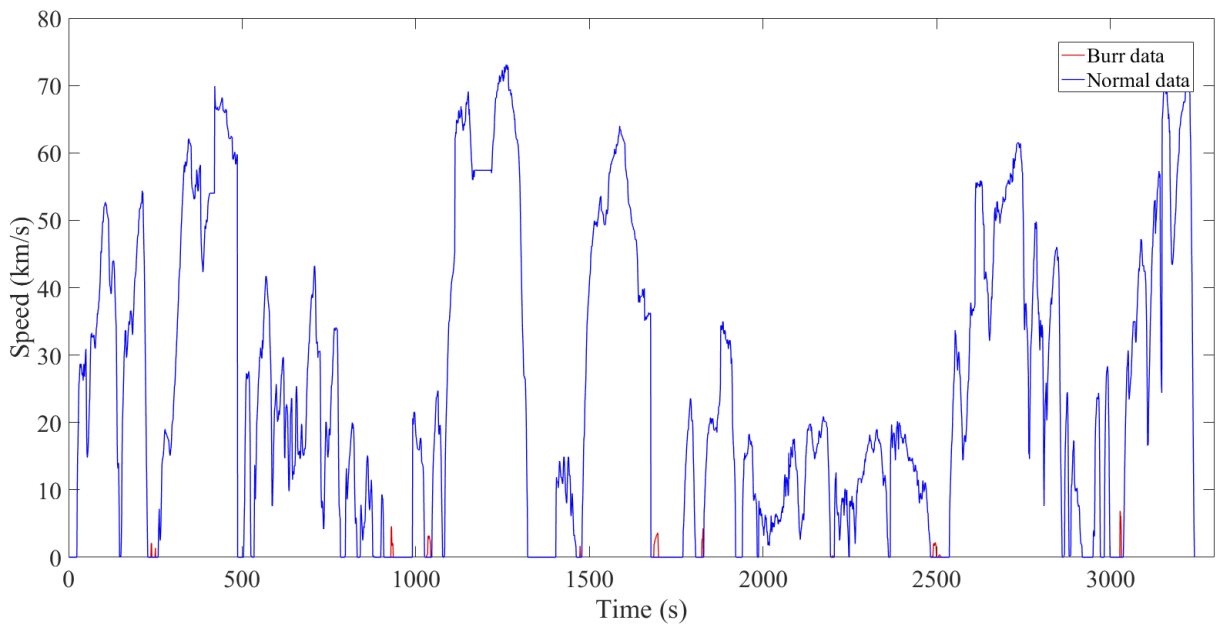

**Fig. 4.** Travel time anomaly data processing

*3.3. Analysis of kinematic sequences*

The kinematic segment refers to the vehicle's driving process from an idle state through a series of acceleration, deceleration, and constant mode to the next idle state. Therefore, a typical kinematics sequence should include four driving modes: idle, acceleration, deceleration, and uniform state. These four states are divided according to the value of acceleration. When it is greater than 0.15 m/s$^2$, named accelerated state; when less than -0.15 m/s$^2$, decelerated state; when less than 0.15 m/s$^2$ and greater than -0.15m/s$^2$, uniform motion; when the velocity zero, idle state. To ensure that the above four driving states are included in one kinematic segment, the cycle length is required to be longer than 20 seconds, and 1701 kinematic segments are finally obtained in combination with the processing of abnormal travel time data during the data processing. In order to analyze the driving characteristics of kinematic segments, 12 kinematic eigenvalues are selected, and the kinematic characteristic calculation results of each kinematic segment using MATLAB are shown in **Table 2**.

**Table 2.**

The calculation results of kinematic sequences parameters

| Number | 1 | 2 | 3 | 4 | … | 788 | 789 | 790 | … | 1698 | 1699 | 1700 | 1701 |
|---|---|---|---|---|---|---|---|---|---|---|---|---|---|
| $T$/s | 67 | 369 | 119 | 218 | … | 37 | 124 | 54 | … | 94 | 248 | 256 | 108 |
| $T_i$/s | 15 | 120 | 25 | 81 | … | 15 | 69 | 30 | … | 42 | 186 | 38 | 28 |
| $T_a$/s | 26 | 99 | 43 | 51 | … | 12 | 23 | 11 | … | 16 | 19 | 61 | 36 |
| $T_d$/s | 17 | 78 | 29 | 35 | … | 8 | 22 | 4 | … | 11 | 13 | 39 | 16 |
| $T_e$/s | 9 | 72 | 22 | 51 | … | 2 | 10 | 9 | … | 25 | 30 | 118 | 28 |
| $S$/km | 0.13 | 2.68 | 0.67 | 1.65 | … | 24.9 | 49.9 | 13.2 | … | 35.3 | 30.7 | 45.3 | 38.7 |
| $V_{max}$/km/h | 17.4 | 56.5 | 45.2 | 55 | … | 7.65 | 14.21 | 3.61 | … | 14.36 | 6.32 | 31.24 | 20.31 |
| $V_m$/km/h | 7.02 | 26.11 | 20.25 | 27.18 | … | 13.21 | 32.29 | 6.84 | … | 25.27 | 22.81 | 34.02 | 26.05 |
| $V_{mr}$/km/h | 7.95 | 35.52 | 25.58 | 41.06 | … | 8.85 | 18.66 | 4.84 | … | 14.05 | 10.82 | 13.34 | 14.34 |
| $V_s$/km/h | 5.24 | 18.5 | 14.25 | 22.82 | … | 0.08 | 0.49 | 0.05 | … | 0.38 | 0.44 | 2.22 | 0.61 |
| $a_a$/m/s$^2$ | 0.42 | 0.41 | 0.48 | 0.42 | … | 0.58 | 0.61 | 0.52 | … | 0.58 | 0.47 | 0.37 | 0.37 |
| $a_d$/m/s$^2$ | -0.64 | -0.53 | -0.72 | -0.64 | … | -0.86 | -0.62 | -1.42 | … | -0.79 | -0.7 | -0.48 | -0.84 |

**4. Methodology**

*4.1. Dimension reduction of parameters*

Dimensionality reduction of characteristic parameters is carried out by PCA [30], and the process is as follows:

*4.1.1. Principal component analysis*

*a) Data standardization*

For the data with the number of sequences of *n* and the parameters of m, the matrix **Z** in $n \times m$ can be obtained as:

$$\mathbf{Z} = \begin{bmatrix} z_{11} & z_{12} & \cdots & z_{1m} \\ z_{21} & z_{22} & \cdots & z_{2m} \\ \vdots & \vdots & \ddots & \vdots \\ z_{n1} & z_{n2} & \cdots & z_{nm} \end{bmatrix} \quad (2)$$

By standardizing *m* variables in **Z**, the mean of the standardized data is zero, and the deviation is one, the standardized matrix **X** is obtained as:

$$\mathbf{X} = \begin{bmatrix} x_{11} & x_{12} & \cdots & x_{1m} \\ x_{21} & x_{22} & \cdots & x_{2m} \\ \vdots & \vdots & \ddots & \vdots \\ x_{n1} & x_{n2} & \cdots & x_{nm} \end{bmatrix} \quad (3)$$

where $x_{ij} = \dfrac{(z_{ij} - \overline{z}_j)}{s_j}$ $i = 1, 2, \cdots, n;\ j = 1, 2, \cdots, m$, $\overline{z}_j = \sum\limits_{i=1}^{n} z_{ij} / n$, and $s_j^2 = \sum\limits_{i=1}^{n} (y_{ij} - \overline{y}_j)^2 / (n-1)$

*b) Calculate the correlation coefficient matrix*

The correlation coefficient matrix **R** calculated as:

$$\mathbf{R} = \begin{bmatrix} r_{11} & r_{12} & \cdots & r_{1m} \\ r_{21} & r_{22} & \cdots & r_{2m} \\ \vdots & \vdots & \ddots & \vdots \\ r_{n1} & r_{n2} & \cdots & r_{nm} \end{bmatrix} \quad (4)$$

where $r_{ij} = \sum\limits_{k=1}^{n}(x_{ki} - \overline{x}_i)(x_{kj} - \overline{x}_j) \bigg/ \sqrt{\sum\limits_{k=1}^{n}(x_{ki} - \overline{x}_i)^2 \sum\limits_{k=1}^{n}(x_{kj} - \overline{x}_j)^2}$.

*c) Calculated eigenvalues $\lambda_i$:*

The magnitude of the eigenvalue represents the contribution of the eigenvector corresponding to the matrix after orthogonalization to the whole matrix. Solve the characteristic equation $|\lambda \mathbf{I} - \mathbf{R}| = 0$. Then arrange them in descending order $(\lambda_1 \geq \lambda_2 \geq \cdots \geq \lambda_m \geq 0)$ and find the orthogonalization

eigenvector $e_i = (i = 1, 2, \cdots, m)$.

*d) Calculate the variance contribution rate and cumulative variance contribution rate*

The variance contribution rate refers to the proportion of a principal component in the total variance. By setting the threshold, the principal component can be extracted to realize the dimensionality reduction of the data. The variance contribution rate and the cumulative variance contribution rate can be calculated as Eq. (5) and Eq. (6):

$$\lambda_i \bigg/ \sum_{k=1}^{m} \lambda_k \quad i = 1, 2, \cdots, m \tag{5}$$

$$\sum_{k=1}^{i} \lambda_i \bigg/ \sum_{k=1}^{m} \lambda_k \quad i = 1, 2, \cdots, m \tag{6}$$

*4.1.2. Principal component selection*

According to the above method and steps, a total of 12 principal components are obtained. The characteristic values and cumulative contribution rates of each principal component are listed in **Table 3**. According to the principle of principal component selection, the cumulative contribution rate is required to be greater than 80%. In addition, the characteristic value of the principal component, as an equally important measurement standard, should be guaranteed to be greater than one. In consideration of these two main principles, the first three principal components are selected by combining the information in **Table 3**. **Table 4** shows the load matrix of the three principal components for the 12 characteristic parameters. The value of each principal component corresponding to the kinematic parameter is its reflection degree to this characteristic. Positive and negative of the value respectively represent the positive and negative correlation between the principal component and the characteristic parameters.

**Table 3**

Principal component variance contribution rate and cumulative variance contribution rate

| Principal component | Eigenvalue | The cumulative variance contribution rate |
|---|---|---|
| 1 | 6.6909 | 0.5576 |
| 2 | 2.2475 | 0.7449 |
| 3 | 1.2797 | 0.8515 |
| 4 | 0.8011 | 0.9183 |
| 5 | 0.5605 | 0.9650 |
| 6 | 0.2155 | 0.9829 |
| 7 | 0.0832 | 0.9899 |
| 8 | 0.0586 | 0.9947 |
| 9 | 0.0279 | 0.9971 |
| 10 | 0.0214 | 0.9989 |
| 11 | 0.0136 | 1.0000 |
| 12 | 4.16E-32 | 1.0000 |

**Table 4**

Principal component load matrix

| Characteristic parameter | Principal component 1 | Principal component 2 | Principal component 3 |
|---|---|---|---|
| Driving time | 0.3325 | 0.2636 | 0.2778 |
| Idle time | 0.0733 | 0.1670 | 0.7666 |
| Acceleration time | 0.3561 | 0.1593 | -0.0224 |
| Decelerate time | 0.3379 | 0.2387 | -0.0077 |
| Uniform time | 0.3289 | 0.2676 | -0.0481 |
| Driving distance | 0.3267 | -0.3266 | 0.0296 |
| Maximum speed | 0.3277 | -0.2288 | -0.2835 |
| Average speed | 0.3216 | -0.3294 | 0.0020 |
| Average travel speed | 0.2707 | -0.4206 | 0.0711 |
| Speed standard deviation | 0.3575 | 0.1404 | -0.0090 |
| Average acceleration | -0.1279 | -0.2065 | 0.4357 |
| Average decelerate | -0.0007 | 0.4918 | -0.2368 |

*4.2. Mean Shift clustering analysis*

*4.2.1. Algorithm theory*

Compared with the K-Means clustering, the Mean Shift algorithm does not need to set the number of clustering in advance and is a sliding window algorithm based on the center of mass. By moving the center point of the region to the mean of the points, the region slides to a denser area, and finally reaches the center point of each category.

For $n$ sample points $x_i$ ($i=1,2,\cdots,n$) in a given dimensional space $R^d$, choose one point $x$, and the primary form of the Mean Shift vector is:

$$M_h(x) = \frac{1}{K} \sum_{x_i \in S_k} (x_i - x) \tag{7}$$

where $S_k$ is the high-dimensional spherical area with radius $h$, defined as:

$$S_h(x) = \left\{ y \mid (y - x_i)(y - x_i)^T \leq h^2 \right\} \tag{8}$$

where $k$ means there are $k$ of the sample points of $x_i$ are in $S_k$.

In the $S_h$ of radius $h$, to reflect the contribution of each sample point $x_i$ to whole sample points, the effect of the offset value on the offset vector can vary with the distance between the sample points and the offset point[27]. The kernel function is added to the primary Mean Shift vector as:

$$M_h(x) = \sum_{i=1}^{n} G(x_i - x/h_i)(x_i - x) \Big/ \sum_{i=1}^{n} G(x_i - x/h_i) \tag{9}$$

where $G(x_i - x/h_i)$ is the kernel function, and the value for the sample points contained in $S_h$.

### 4.2.2. Clustering steps

Based on the spatial distribution of sequences, the Mean shift clustering process is as follows:

**Step1:** Take any 1000 pairs of points from all sample points, calculate the distance between each pair, and take 20% of all distances as radius $h$;

**Step2:** Take an unmarked point from the sample points as the regional center $c$, find out all points where $c$ is the center radius of $h$, denoted as set **M**, consider all points in **M** belonging to cluster **C**, and add the probability of belonging to this class to one for the final classification;

**Step3:** Calculate the vector of each element in **M** to $c$, and add all the vectors to vector **S**;

**Step4:** Move $c$ in the direction of **S**, and the moving distance is $\|\mathbf{S}\|$;

**Step5:** Repeat **Step3, 4**, and **5**, when **S** is **0**, it converges. If the center of the two clusters is less than the threshold, it is recorded as the same type. Otherwise, it increases by one type;

**Step6:** Until all data points are accessed before the iteration ends;

**Step7:** Calculate the frequency of each point being accessed by various types and classify it into the highest frequency.

*4.2.3. Clustering results*

According to the above steps, clustering results of kinematics segments are obtained as shown in **Fig. 5**. The coordinate axis corresponds to the corresponding principal components. The driving cycles represented by the blue points are the first type of driving cycle, which occupy most of the segments and represent the general driving characteristics of the driving data, called the general driving cycles. Besides, the second and third cycles represented by red and green points, which have fewer numbers included and contain some unique driving characteristics, called the special cycles. In addition, there are three segments to have their own category, which are regarded as abnormal cycles.

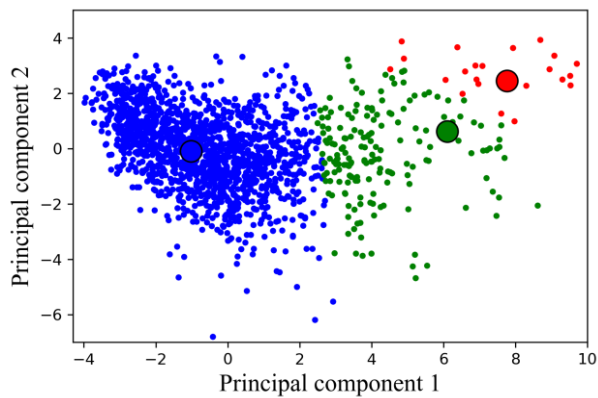
(a)

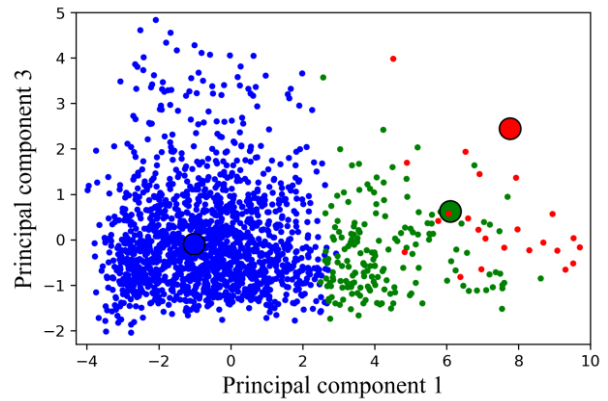
(b)

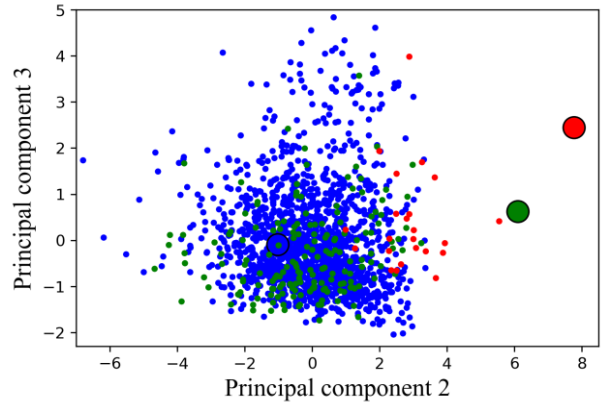

(c)

**Fig. 5.** Clustering results of Mean Shift

**Fig. 6** is the driving cycle curves by randomly selecting segments from the three types of driving cycles. It can be seen that the three cycles have obvious differences in driving time and maximum driving speed. Among them, the segment length of the first condition is the shortest compared with the other two cycles, and the maximum speed is generally less than 50 km/h. The salient feature of the third cycle is that the driving time is significantly longer than the second and third cycle. While the second type of cycle is in the intermediate state between the first the third type of segments in terms of speed and travel time.

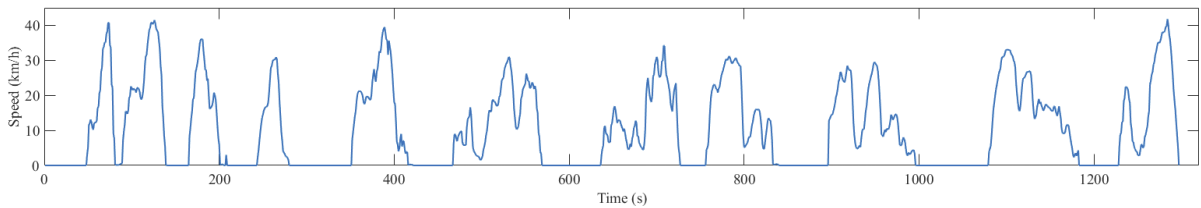

(a)

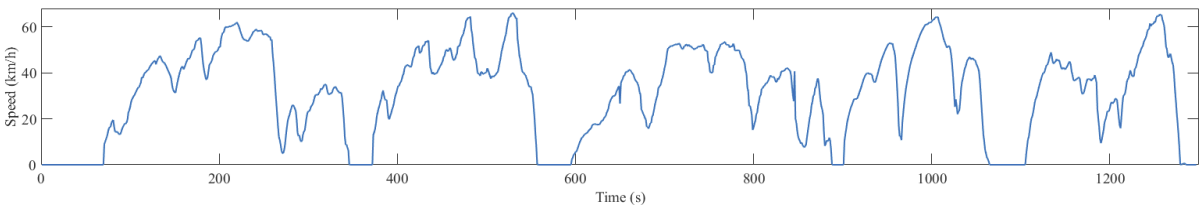

(b)

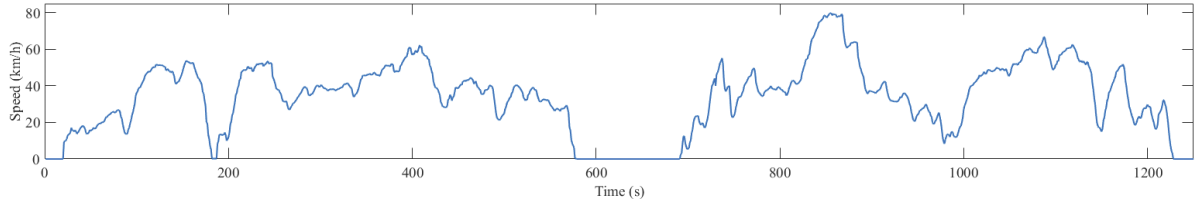

(c)

**Fig. 6.** Curves under the three driving cycles

To further analyze the characteristics of various cycles, the ratios of the idle mode, acceleration, uniform, and deceleration in each cycle were calculated, as shown in **Fig. 7**. It can be seen that the idle state occupies the largest proportion and the uniform state occupies the least proportion in the first type segments. In contrast, to the third segment, the idle state only accounts for about 16%, which is least among the three cycles, but its uniform state is nearly twice of the first segments. While for the four driving states of the second segments were like to be the average level of the three cycles.

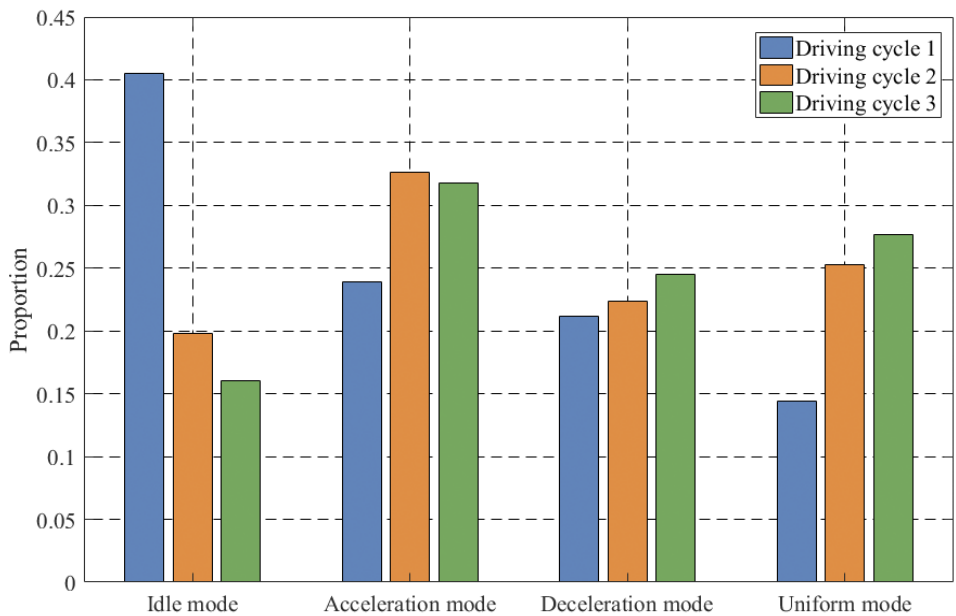

**Fig. 7**. The proportion of driving states under the three driving cycles

## 5. Result and discussion

*5.1. Driving cycles synthetic*

The composition of the driving cycle curve needs to select the representative segments from the three types cycle which can reflect the overall characteristics of its category. Therefore, we selected the average speed, average travel speed, maximum speed, *etc* total of 12 indexes of kinematics characteristic as the selection criteria, and calculate the overall mean for each type in each characteristic. Then figure out the difference between each index in each segment and the overall mean of the type. Furthermore, the index difference is quantified and summed to obtain the overall deviation, and the segments of each category are sorted according to the principle of minimum overall deviation [31].

The final driving cycle curve should cover three type cycles and in consideration of differences in the number of segments among the three categories as the clustering results of the **Section4.2**. The number of segments and travel time in the first cycle (the general cycle) is much larger than the other two kinds of cycles (the special cycles). Therefore, in the synthesis of the final driving cycle curve, one segment of the second and third cycles are respectively selected according to the order of minimum overall deviation, and the rest are selected from the first cycle. The final driving cycle curve is in total 1498seconds as shown in **Fig.8**.

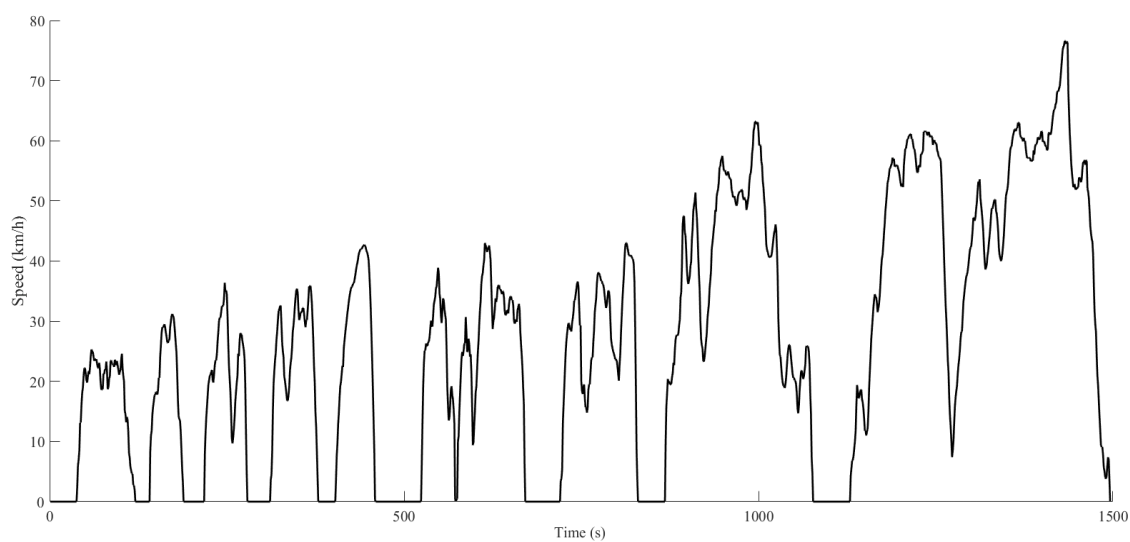

**Fig. 8.** Final driving cycles curve

## 5.2. Discussion

### 5.2.1. Analysis of the driving cycles

**Fig. 9** shows the joint probability distribution of speed and acceleration, in which (a), (b), and (c) correspond to the first, second, and third cycle. By comparison, it can be seen that the curve acceleration probability of the first type of cycle is concentrated at less than -1 m/s$^2$, while the second cycle is concentrated at -1~3 m/s$^2$, and the third cycle is uniformly distributed in each interval. It can be understood that the first type of cycle reflects obvious deceleration characteristics, the second cycle reflects uniform and acceleration characteristics, and the third cycle seems to uniformly reflect each driving state. And (d), (e) in **Fig. 9** respectively correspond to the joint probability distributions of speed-acceleration in the synthetic cycle and the original data, it can be seen that the original driving data is relatively smooth due to the sufficient sample size surface, and the whole probability synthesis cycle can roughly reflect the original speed-acceleration trend, but there are also some differences, for example, compared with the original data the acceleration probability distribution in zero of the synthesis cycle is less apparent. This is related to the small number of segments in the final cycle, and the reliability of this method can be further evaluated by comparing it with other construction schemes.

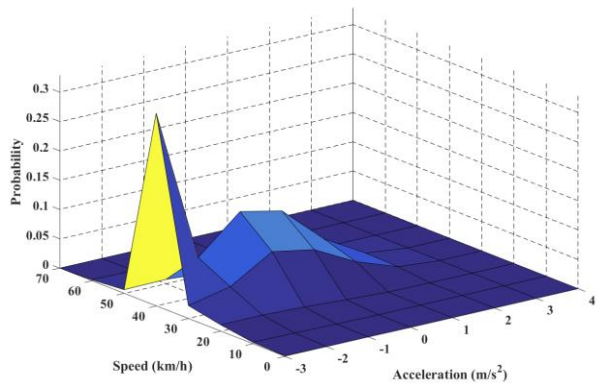

(a)

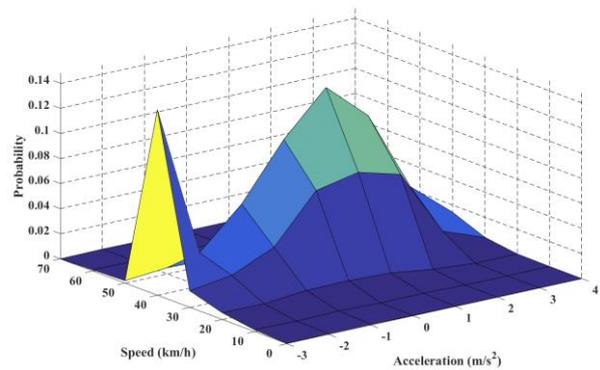

(b)

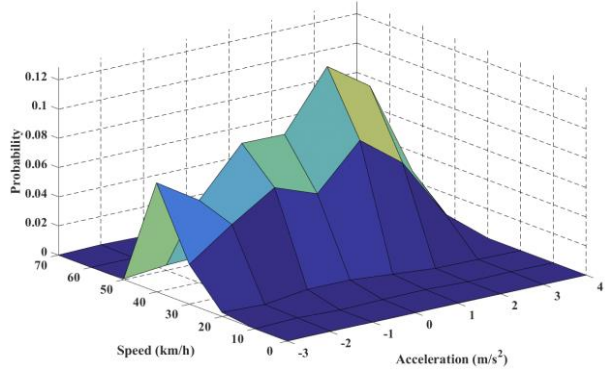 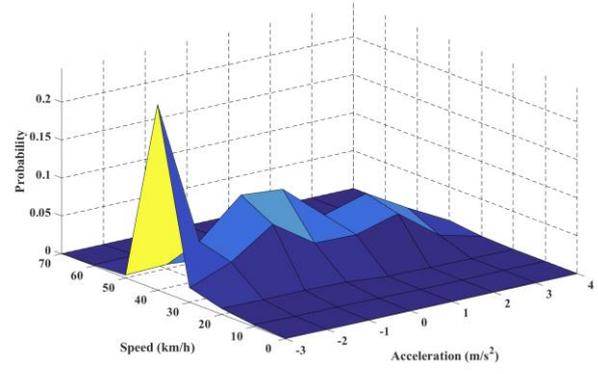

(d) (e)

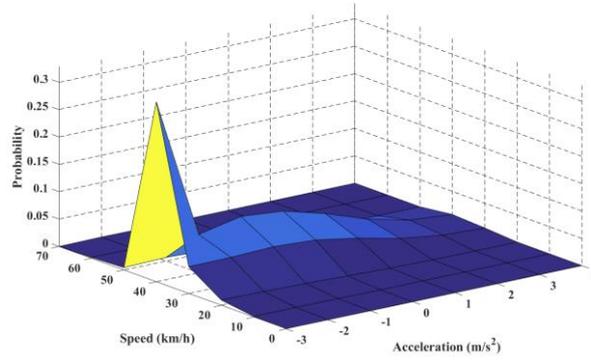

(f)

**Fig. 9.** The probability distribution of speed-acceleration.

In order to verify the effectiveness of this construction method, a comparative analysis was conducted based on the K-Means clustering method [32]. The clustering number was set as three, and we chose eight indicators such as average speed, average travel speed, and acceleration that can reflect the overall characteristics for analysis. **Table 5** is the comparison between the method of this paper and the K-Means construction scheme for the above indicators. It can be seen from **Table 5** that the differences between the two construction schemes in the above 8 indicators are mainly reflected in average speed, average travel speed, and the proportion of each driving mode, while the average acceleration is relatively close. Further, compare the driving cycle constructed by these two methods with the corresponding indexes of the original data to determine which method can better reflect the original data.

**Table 5**

Comparison between K-Means and Mean Shift in driving cycles constriction

| Indicators | First cycle | | Second cycle | | Third cycle | | Driving cycles | |
|---|---|---|---|---|---|---|---|---|
| | K-Means | Mean-Shift | K-Means | Mean-Shift | K-Means | Mean-Shift | K-Means | Mean-Shift |
| $V_m$ | 9.74 | 17.62 | 21.76 | 39.31 | 38.00 | 30.61 | 12.72 | 21.51 |
| $V_{mr}$ | 15.65 | 25.81 | 29.68 | 45.66 | 42.71 | 32.70 | 19.79 | 27.33 |
| $a_a$ | 0.55 | 0.51 | 0.48 | 0.47 | 0.45 | 0.43 | 0.57 | 0.49 |
| $a_d$ | -0.60 | -0.70 | -0.66 | -0.67 | -0.56 | -0.49 | -0.62 | -0.59 |
| $T_i$ | 0.40 | 0.35 | 0.31 | 0.21 | 0.18 | 0.15 | 0.37 | 0.26 |
| $T_a$ | 0.21 | 0.27 | 0.27 | 0.30 | 0.28 | 0.27 | 0.23 | 0.27 |
| $T_d$ | 0.20 | 0.20 | 0.20 | 0.21 | 0.23 | 0.24 | 0.21 | 0.22 |
| $T_e$ | 0.18 | 0.18 | 0.22 | 0.28 | 0.32 | 0.34 | 0.19 | 0.24 |

**Fig. 10** is the value of the corresponding indexes of the real driving cycle, the construction condition based on the K-Means clustering, and the construction condition based on the Mean Shift clustering. **Fig. 11** is the different rate of the corresponding indexes of the two construction methods compared with the real driving data. It can be seen from **Fig. 10** and **Fig. 11** that the Mean Shift clustering construction method is superior to the K-Means method in multiple evaluation parameters, especially in terms of average speed and average travel speed, which is closer to the real driving data.

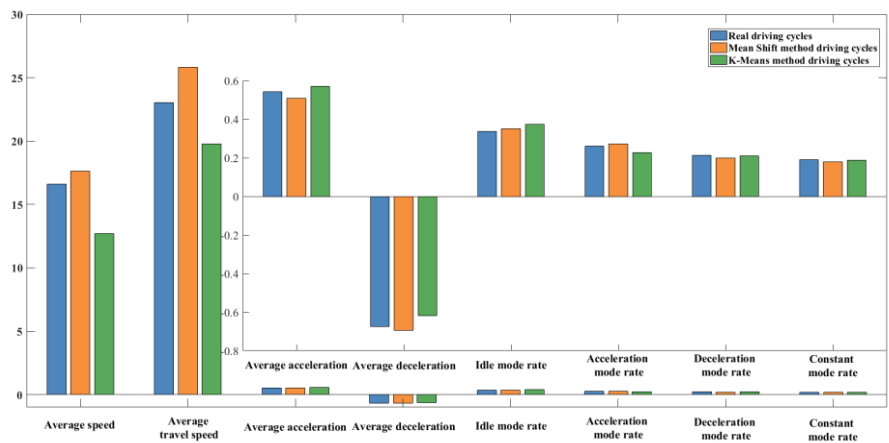

**Fig. 10.** The differences of indicators in the real driving data, the K-Means, and the Mean-Shift method.

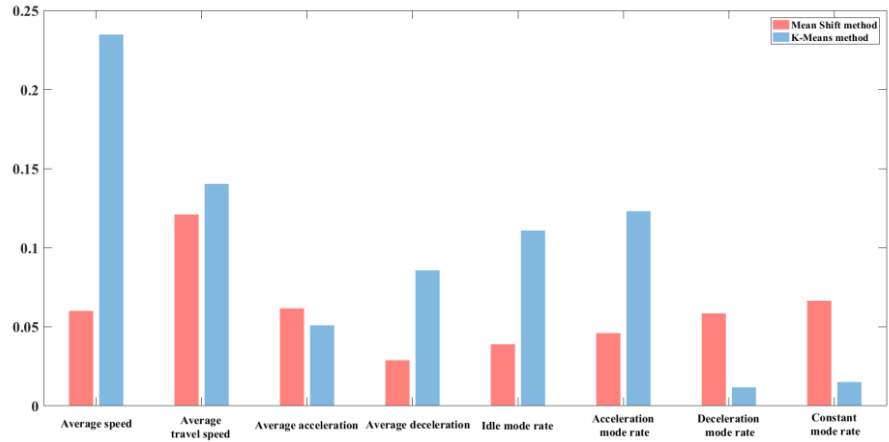

**Fig. 11.** The differences in the error rate of the K-Means and the Mean-Shift driving cycle construction.

By calculating the average difference rate of the driving cycles obtained by these two methods, it is known that the average index difference rate constructed by the Mean Shift clustering method is 6.0%, while the average index difference rate constructed by the K-Means clustering is 9.7%. Therefore, based on the above analysis, the Mean Shift clustering construction method proposed in this paper can better reflect the original driving data.

*5.2.2. Comparison with other countries and cities*

Finally, the final cycle constructed in this paper is also compared with the data of other countries in the world and major cities in China [32-34], to better understand the driving characteristics of vehicles in this city, and the results are shown in **Table 6**. It can be seen that the driving constructed in this paper has some differences in the average speed with the international driving cycles, and it is at the average level in Chinese cities. It is close to Japan's J10-15 driving cycle internationally, but different from the world light vehicle test cycle and NEDC cycle. Compared with Chinese urban light vehicle driving cycles, it is the closest to Beijing's working condition.

**Table 6**

Comparison with other countries and cities

| Driving cycles | $T_i$ | $T_a$ | $T_d$ | $T_e$ | $V_m$ |
| --- | --- | --- | --- | --- | --- |
| Driving cycles in this paper | 0.264 | 0.274 | 0.225 | 0.237 | 21.51 |
| WLTC | 0.127 | 0.309 | 0.286 | 0.278 | 46.42 |
| J10-15 | 0.333 | 0.252 | 0.221 | 0.194 | 22.25 |
| NEDC | 0.226 | 0.232 | 0.166 | 0.375 | 33.34 |
| Beijing | 0.165 | 0.253 | 0.309 | 0.273 | 19.98 |
| Shanghai | 0.316 | 0.228 | 0.233 | 0.223 | 14.96 |
| Guangzhou | 0.178 | 0.291 | 0.272 | 0.260 | 14.14 |
| Shenzhen | 0.202 | 0.323 | 0.291 | 0.184 | 32.39 |

## 6. Conclusion

In this paper, we proposed a method of vehicle driving cycle construction based on the Mean Shift clustering algorithm. Firstly we process the real-world driving data, divide the kinematics sequences, calculate the parameters, and reduce the dimensionality of the parameters through PCA. Then the Mean Shift was used to divide the sequences into three types and select the representative cycles based on the principle of the smallest total deviation of the parameters. Finally, we realized the construction of driving cycles and conducted the further analysis.

Through compares the construct method in this paper with the traditional method of the K-Means clustering algorithm, it is shown that the driving cycle constructed based on the Mean Shift clustering scheme in terms of average speed, average travel speed has more advantages, and the parameters of the overall error is smaller and closer to the vehicles in real road driving conditions. This study implements the innovation of the traditional construction of the micro-trip method. Therefore, the formulation of driving cycle standards according to the method in this paper is of great significance to the research of vehicle energy management and control strategies for future driverless and new-energy vehicles.

By comparing the driving cycles of various countries and cities, it is found that there are large differences. Therefore, when formulating an energy management strategy according to the driving cycles, it is necessary to comprehensively consider the driving characteristics in its own country to make it better adapt to the driving environment.

For further work, we will research the energy management level according to the driving cycle constructed in this paper and use energy consumption as a parameter of the construction process to construct a vehicle cycle incorporating energy efficiency, and eventually, to realize the optimal strategy of vehicle dynamics control and promote the development of new-energy vehicles.

**Founding:**

This work is supported by the National Natural Science Foundation of China (NSFC) (51978310).

**Conflicts of Interest:**

The authors declare no conflict of interest.

**Acknowledgment**

Thanks to all those who provided help and suggested changes during the publication of this article.

**References**

[1] M. Mourad and K. R. Mahmoud, "Performance investigation of passenger vehicle fueled by propanol/gasoline blend according to a city driving cycle," *Energy,* vol. 149, pp. 741-749, 2018.

[2] Y. Huang, N. C. Surawski, B. Organ, J. L. Zhou, O. H. Tang, and E. F. Chan, "Fuel consumption and emissions performance under real driving: Comparison between hybrid and conventional vehicles," *Science of The Total Environment,* vol. 659, pp. 275-282, 2019.

[3] J. Morales-Morales, M. A. Rivera-Cruz, P. Cruz-Alcantar, H. Bautista Santos, I. Cervantes-Camacho, and V. A. Reyes Herrera, "Performance analysis of a hybrid electric vehicle with multiple converter configuration," *Applied Sciences,* vol. 10, no. 3, p. 1074, 2020.

[4] C. Fiori, K. Ahn, and H. A. Rakha, "Power-based electric vehicle energy consumption model: Model development and validation," *Applied Energy,* vol. 168, pp. 257-268, 2016.

[5] P. Lipar, I. Strnad, M. Česnik, and T. Maher, "Development of urban driving cycle with GPS data post processing," *Promet-Traffic&Transportation,* vol. 28, no. 4, pp. 353-364, 2016.

[6] G. Amirjamshidi and M. J. Roorda, "Development of simulated driving cycles for light, medium, and heavy duty trucks: Case of the Toronto Waterfront Area," *Transportation research part D: transport and environment,* vol. 34, pp. 255-266, 2015.

[7] H. He, C. Sun, and X. Zhang, "A method for identification of driving patterns in hybrid electric vehicles based on a LVQ neural network," *Energies,* vol. 5, no. 9, pp. 3363-3380, 2012.

[8] P. Yuhui, Z. Yuan, and Y. Huibao, "Development of a representative driving cycle for urban buses based on the K-means cluster method," *Cluster Computing,* vol. 22, no. 3, pp. 6871-6880, 2019.

[9] S. Shi *et al.*, "Research on Markov property analysis of driving cycles and its application," *Transportation Research Part D: Transport and Environment,* vol. 47, pp. 171-181, 2016.

[10] M. Zhang, S. Shi, N. Lin, and B. Yue, "High-efficiency driving cycle generation using a Markov chain evolution algorithm," *IEEE Transactions on Vehicular Technology,* vol. 68, no. 2, pp. 1288-1301, 2018.

[11] D.-y. Zheng, X.-g. Wu, H. Chen, and J.-y. Du, "Construction of driving conditions of Harbin urban passenger cars," *Journal of highway and transportation research and development (English edition),* vol. 12, no. 1, pp. 81-88, 2018.

[12] I. T. Jolliffe and J. Cadima, "Principal component analysis: a review and recent developments," *Philosophical Transactions of the Royal Society A: Mathematical, Physical and Engineering Sciences,* vol. 374, no. 2065, p. 20150202, 2016.

[13] H. Tong, "Development of a driving cycle for a supercapacitor electric bus route in Hong Kong," *Sustainable Cities and Society,* vol. 48, p. 101588, 2019.

[14] L. D. Ragione and G. Meccariello, "The evaluation of a new kinematic emissions model on real and simulated driving cycles," *SAE International Journal of Fuels and Lubricants,* vol. 3, no. 2, pp. 521-531, 2010.

[15] K. Kivekäs, J. Vepsäläinen, and K. Tammi, "Stochastic driving cycle synthesis for analyzing the energy consumption of a battery electric bus," *IEEE Access,* vol. 6, pp. 55586-55598, 2018.

[16] Q. Shi, B. Liu, Q. Guan, L. He, and D. Qiu, "A genetic ant colony algorithm-based driving cycle generation approach for testing driving range of battery electric vehicle," *Advances in Mechanical Engineering,* vol. 12, no. 1, p. 1687814019901054, 2020.

[17] A. Fotouhi and M. Montazeri-Gh, "Tehran driving cycle development using the k-means clustering method," *Scientia Iranica,* vol. 20, no. 2, pp. 286-293, 2013.

[18] D. Förster, R. B. Inderka, and F. Gauterin, "Data-Driven Identification of Characteristic Real-Driving Cycles Based on k-Means Clustering and Mixed-Integer Optimization," *IEEE Transactions on Vehicular Technology,* vol. 69, no. 3, pp. 2398-2410, 2019.

[19] Z. Chen, C. Yang, and S. Fang, "A Convolutional Neural Network-Based Driving Cycle Prediction Method for Plug-in Hybrid Electric Vehicles With Bus Route," *IEEE Access,* vol. 8, pp. 3255-3264, 2019.

[20] X. Zhao, Q. Yu, J. Ma, Y. Wu, M. Yu, and Y. Ye, "Development of a representative EV urban driving cycle based on a k-means and SVM hybrid clustering algorithm," *Journal of Advanced*

*Transportation,* vol. 2018, 2018.
[21] M. E. Celebi, H. A. Kingravi, and P. A. Vela, "A comparative study of efficient initialization methods for the k-means clustering algorithm," *Expert systems with applications,* vol. 40, no. 1, pp. 200-210, 2013.
[22] S. E. Y. Nayini, S. Geravand, and A. Maroosi, "A novel threshold-based clustering method to solve K-means weaknesses," in *2017 International Conference on Energy, Communication, Data Analytics and Soft Computing (ICECDS)*, 2017, pp. 47-52: IEEE.
[23] S.-J. Chang-Chien, W.-L. Hung, and M.-S. Yang, "On mean shift-based clustering for circular data," *Soft Computing,* vol. 16, no. 6, pp. 1043-1060, 2012.
[24] O. Hyrien and A. Baran, "Fast nonparametric density-based clustering of large datasets using a stochastic approximation mean-shift algorithm," *Journal of Computational and Graphical Statistics,* vol. 25, no. 3, pp. 899-916, 2016.
[25] S.-Y. Shiu and T.-L. Chen, "On the strengths of the self-updating process clustering algorithm," *Journal of Statistical Computation and Simulation,* vol. 86, no. 5, pp. 1010-1031, 2016.
[26] X. Wang, W. Qiu, and J. Wu, "Convergence and stability analysis of mean-shift algorithm on large data sets," *Statistics and Its Interface,* vol. 9, no. 2, pp. 159-170, 2016.
[27] R. Yamasaki and T. Tanaka, "Properties of mean shift," *IEEE transactions on pattern analysis and machine intelligence,* 2019.
[28] J. Brady and M. O'Mahony, "Development of a driving cycle to evaluate the energy economy of electric vehicles in urban areas," *Applied energy,* vol. 177, pp. 165-178, 2016.
[29] S.-H. Ho, Y.-D. Wong, and V. W.-C. Chang, "Developing Singapore Driving Cycle for passenger cars to estimate fuel consumption and vehicular emissions," *Atmospheric environment,* vol. 97, pp. 353-362, 2014.
[30] H. Abdi and L. J. Williams, "Principal component analysis," *Wiley interdisciplinary reviews: computational statistics,* vol. 2, no. 4, pp. 433-459, 2010.
[31] J. Zhang, Z. Wang, P. Liu, Z. Zhang, X. Li, and C. Qu, "Driving cycles construction for electric vehicles considering road environment: A case study in Beijing," *Applied Energy,* vol. 253, p. 113514, 2019.
[32] Z. Liu, P. Zhu, X. Liu , and Z. Liu, "Research on Improved K-Means and Driving Cycle Construction," *Automobile Technology,* vol. 11, pp. 57-62, 2019.
[33] h. Zhang, Y. Yao, and X. Yang, "Light-duty Vehicles Driving Cycle Construction Based on Urban Roads," *Journal of Southwest Jiaotong University,* vol. 54, no. 6, pp. 1139-1146+1154, 2018.
[34] Y. Peng, H. Yang, M. Li, and X. Qiao, "Research on the Construction Method of Driving Cycle for the City Car Based on K-Means Cluster Analysis," *Automobile Technology,* vol. 11, pp. 13-18, 2017.